\PassOptionsToPackage{table, dvipsnames}{xcolor}

\documentclass[sigconf, authorversion, noacm]{acmart}

\usepackage{mystyle}

\def\doctitle{Continuous Deep Learning:\\A Workflow to Bring Models into Production}

\def\docauthors{Janosch Baltensperger, Pasquale Salza, Harald C. Gall}

\def\dockeywords{%
Deep learning, machine learning, continuous integration, software maintenance and evolution
}

\StrSubstitute{\doctitle}{\\}{ }[\cleandoctitle]
\StrSubstitute{\dockeywords}{.}{}[\cleandockeywords]
\hypersetup{
	pdftitle={\cleandoctitle},
	pdfauthor={\docauthors},
	pdfkeywords={\cleandockeywords}
}

\def\replicationurl{https://github.com/janousy/CDL}

\DeclareAcronym{ml}{
	short = ML,
	long = {machine learning}
}

\DeclareAcronym{dl}{
	short = DL,
	long = {deep learning}
}

\DeclareAcronym{nn}{
	short = NN,
	long = {neural network}
}

\DeclareAcronym{ai}{
	short = AI,
	long = {artificial intelligence}
}

\DeclareAcronym{ci}{
	short = CI,
	long = {continuous integration}
}

\DeclareAcronym{cd}{
	short = CD,
	long = {continuous delivery}
}

\DeclareAcronym{gpu}{
	short = GPU,
	long = {graphics processing unit}
}

\DeclareAcronym{cpu}{
	short = CPU,
	long = {central processing unit}
}

\DeclareAcronym{vcpu}{
	short = vCPU,
	long = {virtual central processing unit}
}

\DeclareAcronym{api}{
	short = API,
	long = {application programming interface}
}

\DeclareAcronym{uzh}{
	short = UZH,
	long = {University of Zurich}
}

\DeclareAcronym{vm}{
	short = VM,
	long = {virtual machine}
}

\DeclareAcronym{ram}{
	short = RAM,
	long = {random-access memory}
}

\DeclareAcronym{vcs}{
	short = VCS,
	long = {version control system}
}

\DeclareAcronym{ui}{
	short = UI,
	long = {user interface}
}

\DeclareAcronym{yaml}{
	short = YAML,
	long = {yet another markup language}
}

\DeclareAcronym{cli}{
	short = CLI,
	long = {command-line interface}
}

\DeclareAcronym{nlp}{
	short = NLP,
	long = {natural language processing}
}

\DeclareAcronym{json}{
	short = JSON,
	long = {JavaScript object notation}
}

\DeclareAcronym{rest}{
	short = REST,
	long = {representational state transfer}
}

\DeclareAcronym{asha}{
	short = ASHA,
	long = {asynchronous successive halving algorithm}
}

\DeclareAcronym{hpc}{
	short = HPC,
	long = {high performance computing}
}

\DeclareAcronym{hp}{
	short = HP,
	long = {hyperparameter}
}

\begin{document}

\title[\cleandoctitle]{\doctitle}

\author{Janosch Baltensperger}
% \orcid{0000-0000-0000-0000}
\affiliation{%
	\institution{University of Zurich}
	\country{Switzerland}
}
\email{janosch.baltensperger@uzh.ch}

\author{Pasquale Salza}
\orcid{0000-0002-8687-052X}
\affiliation{%
	\institution{University of Zurich}
	\country{Switzerland}
}
\email{salza@ifi.uzh.ch}

\author{Harald C. Gall}
\orcid{0000-0002-3874-5628}
\affiliation{%
	\institution{University of Zurich}
	\country{Switzerland}
}
\email{gall@ifi.uzh.ch}

\renewcommand{\shortauthors}{Baltensperger et al.}

\begin{abstract}

% Deep learning has gained immense attraction with the emergence of big data and advanced computing power.
% Through the use of artificial neural networks, various breakthroughs were achieved in fields such as language understanding and image recognition.
% Nevertheless, it has soon become clear that deep learning and machine learning in general impose various additional challenges besides building an accurate model.
Researchers have been highly active to investigate the classical machine learning workflow and integrate best practices from the software engineering lifecycle.
However, deep learning exhibits deviations that are not yet covered in this conceptual development process.
This includes the requirement of dedicated hardware, dispensable feature engineering, extensive hyperparameter optimization, large-scale data management, and model compression to reduce size and inference latency.
Individual problems of deep learning are under thorough examination, and numerous concepts and implementations have gained traction.
Unfortunately, the complete end-to-end development process still remains unspecified.
In this paper, we define a detailed deep learning workflow that incorporates the aforementioned characteristics on the baseline of the classical machine learning workflow.
We further transferred the conceptual idea into practice by building a prototypic deep learning system using some of the latest technologies on the market.
To examine the feasibility of the workflow, two use cases are applied to the prototype.
% The first use case represented a text classification problem, while the second use case focused on image processing.
% We thereby successfully demonstrated the application of the workflow on distinct examples.
% In summary, it becomes apparent that the deep learning lifecycle compromises a large set of steps and involves various roles.
% With our defined workflow, we present a profound guideline for the deep learning development process.
% Moreover, we conclude that the technologies currently available on the market are not fully mature.
% Great effort is required to manage all deep learning artifacts and keep versions aligned within continuous iterations over the lifecycle.

\end{abstract}




% Hacks for author version
\setcopyright{none}
\settopmatter{printacmref=false}
\renewcommand\footnotetextcopyrightpermission[1]{}
\settopmatter{printfolios=true}
\pagestyle{plain}

\maketitle

\section{Introduction}
\label{sec:introduction}

The traditional software engineering lifecycle is usually maintained through \ac{ci} and \ac{cd} to enable planning, development, and deployment of a software artifact.
Moreover, DevOps, which stands for development and operations, has manifested as a practice (and even as a culture) to merge continuous lifecycle management into a single set of processes.
% \todo{Pasquale}{Add citation.}
Although many software engineering principles can be transferred to \ac{ml} development, various new challenges have to be solved~\cite{lorenzoni_machine_2021}.
In comparison, \ac{ml} projects exhibit many critical differences.
For example, development is data-centric instead of code-centric, individual modules are hard to isolate, and the project team is more diverse in terms of the required skill-set~\cite{amershi_software_2019}.
Maintenance is more difficult and costly than development, as a model needs to be continuously improved and adapted to a changing environment~\cite{sculley_hidden_2015}.
This leads to more frequent iterations over the workflow compared to classical software engineering.
Due to the non-deterministic behavior, \ac{ml} becomes a highly experimental process, which brings up the need for reproducibility~\cite{zaharia_accelerating_2018}.

%Although the \ac{ml} lifecycle is different from traditional software development~\cite{amershi_software_2019}, %a similar framework named MLOps has been introduced, \ie, \ac{ml} operations.
%In this process setting, data scientists create the development environment, while software/operation %engineers are responsible for the setup of the production environment~\cite{Karamitsos2020}.

In general, \ac{ml} lifecycles require sophisticated pipelines, which facilitate data management, training, deployment, and model integration into the corresponding product.
A previous case study at Microsoft~\cite{amershi_software_2019} was conducted to describe the current concept of \ac{ml} development and proposed an abstract workflow reaching from model requirements up to model monitoring in production.
Principally, a workflow consists of an ordered sequence of activities required to achieve one or multiple goals~\cite{aalst_workflow_2002}.
An activity is defined as a task contributing to the defined objective, which can be performed either automatically or manually by a determined individual.
Regarding information technology, a workflow provides systematic organization and reproducibility to the development process,
while reducing costs and increasing productivity~\cite{asuman_workflow_1998}.

% \paragraph{Problem domain}
In the context of \ac{dl}, there is still a lack of guidance when it comes to integrating it into the software development process.
Such a workflow may deviate from the one proposed by Microsoft~\cite{amershi_software_2019} for multiple reasons:
\begin{inparaenum}[(1)]
    \item while conventional \ac{ml} is based on manual feature engineering, \ac{dl} is based on an end-to-end approach, \ie, features are learned automatically~\cite{miao_unified_2017};
    \item a large amount of high-quality data is required to accurately learn data representations, which introduces the need for scalable data management pipelines;
    \item as \acp{nn} are computationally intensive, specialized hardware is required. This includes the use of \acp{gpu} for parallelization and distributed training ~\cite{aida-zade_comparison_2019};
    \item \ac{dl} requires the configuration of numerous \acp{hp}, which need to be optimized;
    \item trained \acp{nn} tend to be large and resource-intensive, introducing the need for model compression, \eg, pruning and quantization, low-rank factorization, convolutional filters, or knowledge distillation~\cite{cheng_survey_2020}.
    % This is especially the case for deployment environments with limited computing power~\cite{Polino2018}, or low-latency real-time serving~\cite{Hazelwood2018}.
\end{inparaenum}
% Moreover, high performance graphical processing units (\ac{gpu}) are recommended to be able to train a neural network in reasonable time, since the training process is computationally expensive~\cite{aida-zade_comparison_2019}.

Several of these critical aspects of \ac{dl} have been addressed independently in past research.
For instance, advanced algorithms have been proposed to accelerate the process of finding the optimal parameters, and new platforms have been introduced to make \ac{gpu} training more accessible to researchers and practitioners~\cite{hutter_automated_2019, nvidia_inference}.
However, the concepts and technologies presented are still relatively new and not yet fully mature~\cite{haakman_ai_2021}.
Additionally, these individual solutions have not yet been assembled to an end-to-end development process for \ac{dl}.
A conceptual representation of the complete workflow and a corresponding implementation remains undefined.

% \paragraph{Contribution}
To overcome the above-stated knowledge gap, this paper aims at investigating the current development workflow of \ac{dl} in the context of \ac{ml} lifecycle management.
The goal is to specify best-practice guidelines from model development to deployment and execution, \ie, bringing \ac{dl} models into production.
Therefore, lifecycle critical differences to conventional \ac{ml} are collected and integrated into an extended workflow for the \ac{dl} lifecycle.
Additionally, a prototype is implemented to demonstrate the practicability of the defined workflow.
Overall, this paper summarizes the current state of research focused on the \ac{dl} workflow and investigates its applicability in practice.
Then, we demonstrate that our abstract definition of the workflow can be utilized in common \ac{dl} applications.
% Moreover, despite the construction of an effective prototype, we indicate that the technologies currently available on the market are not fully mature.
The technical instructions on the prototype and source code for each of the use cases are published at the address \url{\replicationurl}.

\smallskip
\noindent
% \paragraph{Paper organization}
This paper is organized as follows.
% ~\cref{sec:background} summarizes the theoretical baseline for this paper, which includes the workflow defined for classical \ac{ml}, the roles involved in a \ac{ml} team and the decisive characteristics of \ac{dl}.
\Cref{sec:related_work} describes the correlated research.
We derive the abstract \ac{dl} workflow in~\cref{sec:workflow}, using the collected particularities of \ac{dl}.
This abstract definition is then implemented in a minimum viable prototype in \cref{sec:prototype_implementation}, where we select a set of open source technologies to facilitate end-to-end \ac{dl}.
The usage of our prototype and consequently the applicability of our abstract workflow is then shown through two distinct use cases in \cref{sec:use_cases}, based on a text classification and image processing problem.
Our findings are ultimately concluded in \cref{sec:conclusions}.

\section{Related Work}
\label{sec:related_work}

While algorithms and frameworks to build \ac{ml} models evolved quickly, other stages of the workflow have been neglected for a long time.
However, to integrate \ac{ml} into the current software applications, a need for a conceptual development process emerged.
In this section, we describe the work related to our research.

\paragraph{\Acl{ml} workflow}
Amershi \etal~\cite{amershi_software_2019} investigated the development of \ac{ml} applications at Microsoft.
Through a case study, a high-level concept of the \ac{ml} workflow composed of nine steps was deduced.
Furthermore, they highlighted the current challenges imposed during the \ac{ml} development and introduced a model to measure the \ac{ml} process maturity.

Later, Salama \etal~\cite{google_mlops} took the \ac{ml} workflow a step further.
From a more practical perspective, they presented a conceptual representation of a fully integrated \ac{ml} system targeted towards continuous adaption to the business environment.
Within their work, it is illustrated what artifacts are produced during the workflow and how data is moved and transformed between stages.

Compatibly, Garcia \etal~\cite{garcia_context_2018} argue that a crucial piece currently missing in the \ac{ml} workflow was the context.
Unstructured and offhand transitions between stages, and consequently between roles, could impair productivity and reproducibility.
Therefore, artifacts produced by an individual role should not appear as a black-box to other team members.

Furthermore, within the work of Haakman \etal~\cite{haakman_ai_2021}, it is argued that several steps within the \ac{ml} lifecycle had been neglected up to now.
The authors interviewed \num{17} \ac{ml} practitioners at ING, a company which operates in the fintech industry.
These interviews revealed that many existing workflow models do not compromise crucial steps such as data collection, feasibility study, documentation, risk assessment, model evaluation and monitoring.
They stress that the \ac{ml} development process should not only focus on algorithms, but the complete lifecycle.
Additionally, it is stated that the existing tools for \ac{ml} were not mature enough.
Many practitioners would still rely on manual solutions, despite the existence of automating technologies.
Indeed, there is a broad set of tools available on the market, with many still being in their early development phases~\cite{mlops, mlops_community}.
Moreover, very few specifically target \ac{dl}.

\paragraph{\Acl{dl} lifecycle}
Miao \etal~\cite{miao_unified_2017} addressed the issue of managing models and their corresponding artifacts.
They built a lifecycle management system for versioning models and a domain-specific language to query created \ac{dl} models.
Thereby, users can explore and compare hyperparameter tuning experiments using external frameworks and publish models.

Instead, Zhang \etal~\cite{zhang_empirical_2019} conducted an empirical study concerning common challenges within \ac{dl} development.
By collecting questions and answers on \textit{Stack Overflow} and building a classification model, they concluded five categories of common issues: \ac{api} misuse, incorrect hyperparameter selection, \ac{gpu} computation, static graph computation and limited debugging and profiling support.
It is further stated that the current tool chain was not fully mature.

% Moreover, Guo \etal~\cite{Guo2019} examined the influence of platforms and frameworks on \ac{dl} development.
% Findings showed that a model suffers from diminishing accuracy when converting to another \ac{dl} framework.
% They resulted to the conclusion that there would be a demand for a universal \ac{dl} platform.
% Additionally, various practical guidelines are presented regarding stability and robustness of existing frameworks.

\smallskip
\noindent
In summary, we argue that no investigation has yet been conducted on the complete workflow specifically for \ac{dl}.
Most research focuses on either a high-level abstraction of classical \ac{ml} or the implementation of specific stages of the \ac{dl} workflow.

\section{A Deep Learning Workflow}
\label{sec:workflow}

In this paper, we want to deduce the workflow required to perform end-to-end \ac{dl} and demonstrate the usage of the workflow through a minimum viable prototype.
Therefore, this section outlines the abstract \ac{dl} workflow that complies with the requirements listed in \cref{sec:introduction}.
First, we give a high-level overview of the components and roles required.
Then, we provide a more detailed description of all activities and interactions.

\subsection{High Level Overview}
\label{subsec:overview_flow}

\begin{figure}[tb]
	\centering
	\includegraphics[width=0.99\linewidth]{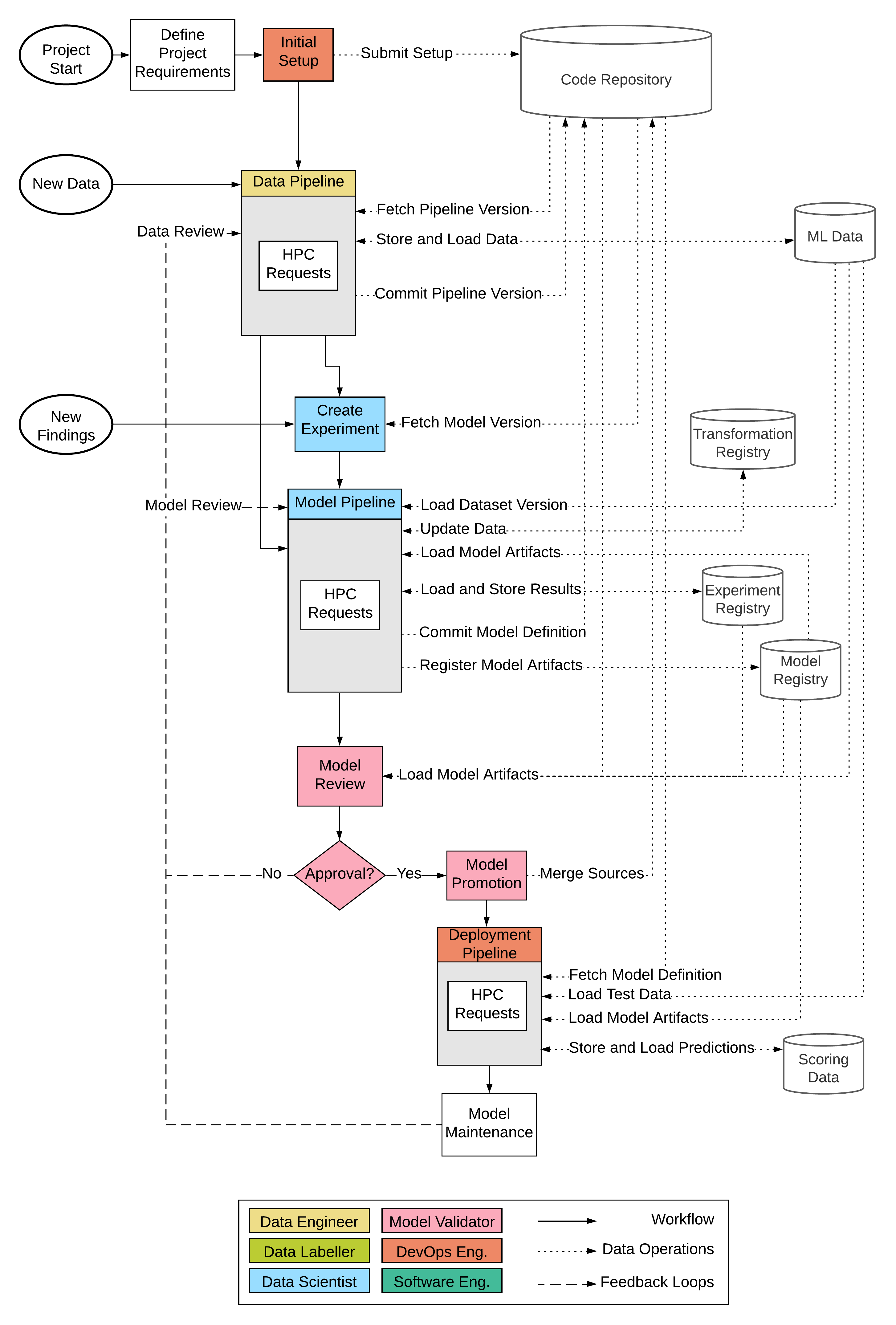}
	\caption{High-level overview of the \ac{dl} workflow.}
	\label{fig:overview_flow}
\end{figure}

We describe the \ac{dl} workflow as a flow chart, which defines the order of all activities and the corresponding roles to take responsibility.
Furthermore, the flow chart demonstrates what data and operations may be involved at each step.
The complete overview flow chart is visualized in \cref{fig:overview_flow}.
It is important to mention that some steps described in this abstract workflow can be optional.
We define the responsibility based on the general definition of a role in a data science team.
% However, the allocation mostly depends on the project setting, and the roles responsible for a specific activity are therefore not fixed.

Within the workflow, we distinguish between \stepNeutral{workflow steps} and \entityNeutral{persistence entities}, which store data produced by a workflow step, once an \actorNeutral{actor} performed an action.
We define the following types of \entityNeutral{persistence entities}:
\begin{defs}
	\item [\entityNeutral{Code Repository}] stores source code within version control and allows sharing.
	\item [\entityNeutral{ML Data}] holds versioned testing and training data prepared by a \actorDataEngineer{}, including the corresponding metadata and labels.
	\item [\entityNeutral{Transformation Registry}] stores preprocessed data specific to a model, produced by a \actorDataScientist{} for faster and more convenient access.
	\item [\entityNeutral{Experiment Registry}] tracks configuration, metrics, and results from an experiment conducted by a \actorDataScientist{}.
	\item [\entityNeutral{Model Registry}] holds model artifacts of all models registered.
	This includes model definition (source code), configuration (\acp{hp}, environment, etc.), metadata (version, creator, time, etc.), dependencies (\eg, software packages, files), and most importantly the serialized model, \ie,  the trained weights in binary format.
	\item [\entityNeutral{Scoring Data}] stores prediction requests and results for analysis and monitoring.
\end{defs}

Our workflow initially only allows one starting point, namely the \stepNeutral{Project Start}.
Moreover, there is no point of termination as \ac{dl}, as well as classical \ac{ml}, is a cyclic process due to continuous model improvements and adaptions to a possibly changing environment.
Upon further iterations of the lifecycle, various roles have the option to step in at almost any stage of the workflow, either due to feedback loops or individual initiatives.

At the beginning of every \ac{dl} project, the requirements need to be defined.
This is usually an interdisciplinary activity, involving business, research, and engineering~\cite{mark_introducing_2020}.
After coming to an agreement, \ie, \stepNeutral{Define Project Requirements}, a \actorDevOpsEngineer{} is responsible for the \stepDevOpsEngineer{Initial Setup}.
They introduce and maintain tools, methodologies and processes for the team to support development, deployment, and monitoring of \ac{ml} models~\cite{redhat_devops}.
This includes the version-controlled \entityNeutral{Code Repository} and other technical infrastructure.
The \entityNeutral{Code Repository} represents a central location to store, version, and share source code, so that the complete workflow remains reproducible.
Once the setup is complete, other team members can start with their implementations.

Theoretically, the \ac{dl} workflow comprehends three fundamental pipelines, namely the \stepDataEngineer{Data Pipeline}, \stepDataScientist{Model Pipeline}, and \stepDevOpsEngineer{Deployment Pipeline}, which are explicitly discussed in their respective sections.
All pipelines share the requirement of \ac{hpc} Requests, involve a subset of roles, and interact with the defined \entityNeutral{persistence entities}.
These pipelines are not necessarily tied to a specific order on a linear timeline and can be executed independently.
Nevertheless, the objective is to derive to a model in production.
Once a model is deployed for inference, the final step of a workflow iteration, the \stepNeutral{Model Maintenance}, is determining when and where to re-enter the workflow.
This can occur at various stages, illustrated by feedback loops on the flow chart.

\subsection{Data Pipeline}
\label{subsec:data_flow}

\begin{figure}[tb]
	\centering
	\includegraphics[width=0.8\linewidth]{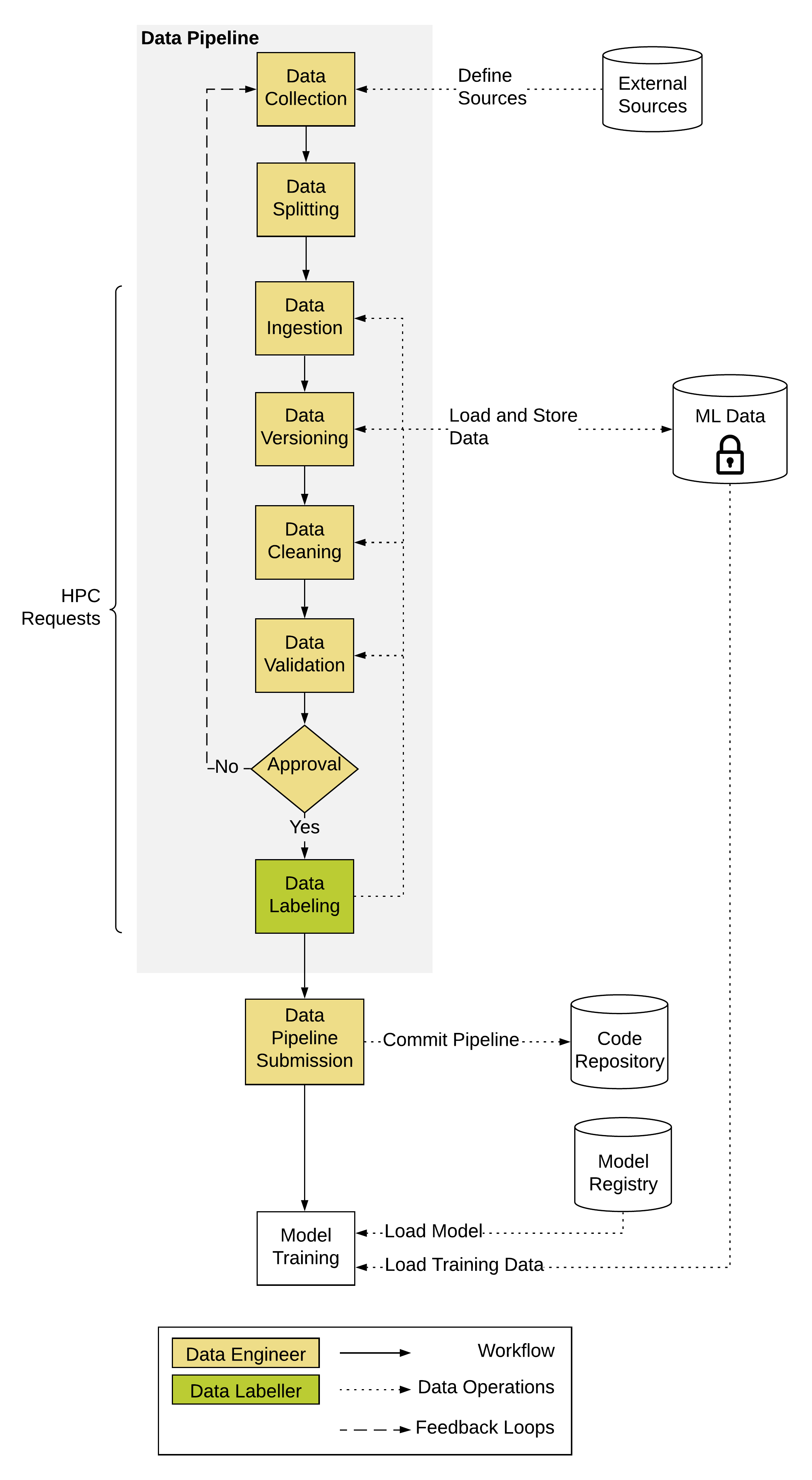}
	\caption{Definition of the abstract data pipeline.}
	\label{fig:data_flow}
\end{figure}

% \begin{figure*}[th]
% 	\centering
% 	\includegraphics[width=1.0\linewidth]{figures/workflow/abstract_data_v2}
% 	\caption{Definition of the abstract data pipeline.}
% 	\label{fig:data_flow}
% \end{figure*}

The data pipeline, displayed as a flow chart in \cref{fig:data_flow}, is the fundamental element of the \ac{dl} workflow, as a \ac{dl} model directly depends on the supplied data~\cite{munappy_data_2019}.
There are essentially two roles present within the data pipeline: the \actorDataEngineer{} and the \actorDataLabeler{}.
A \actorDataEngineer{}'s responsibility is to construct, acquire, prepare, and store data securely~\cite{lakshmanan_machine_2020}.
A \actorDataLabeler{}, on the other hand. provides an accurate ground-truth for supervised or semi-supervised learning.

The first step in the data pipeline, \ie, \stepDataEngineer{Data Collection}, is defining the external sources for the data.
If the test and training sets do not have different origins, the data engineer divides the collected data in a reproducible manner at the subsequent step of \stepDataEngineer{Data Splitting}.
This step may not be required for further iterations if the two datasets are updated independently.
Once the sources for the training and test data are defined, the \stepDataEngineer{Data Ingestion} step is targeted towards loading the external data into a suitable storage option, illustrated as \entityNeutral{ML Data} in \cref{fig:data_flow}.
\stepDataEngineer{Data Versioning} is indispensable and critical for data lineage~\cite{lakshmanan_machine_2020}, and thus performed directly after \stepDataEngineer{Data Ingestion}, presumably in an automated fashion.
Afterwards, the \actorDataEngineer{} defines how the ingested data is to be \stepDataEngineer{cleaned} and \stepDataEngineer{validated}, before deciding whether the amount and quality of the data is sufficient for training a \ac{dl} model.
If this is not the case, the data engineer returns to the step of \stepDataEngineer{Data Collection}.
Otherwise, the prepared datasets are approved and released for \stepDataLabeler{Data Labeling}, by the internal or external \actorDataLabeler{}.

The datasets generated by the data pipeline are required to be reproducible.
Thus, not only the datasets themselves require versioning, but also the process that produced these datasets.
The \actorDataEngineer{} thus commits the definitions of all steps performed within the data pipeline to the version controlled \entityNeutral{Code Repository}, initially set up by the \actorDevOpsEngineer{}.
On further iterations of the \ac{dl} lifecycle, when there is already a model available, one can optionally trigger the execution of \stepNeutral{Model Training} to analyze the performance with the new or updated dataset as a form of \textit{Continuous Integration}.

When working with large and complex data, various steps within the data pipeline such as \stepDataEngineer{Data Ingestion}, \stepDataEngineer{Data Cleaning}, and \stepDataEngineer{Data Validation} are resource demanding.
Thus, a \actorDataEngineer{} should request computing resources on-demand within the steps of the pipeline.

\subsection{Model Pipeline}
\label{subsec:model_flow}

\begin{figure}[tb]
	\centering
	\includegraphics[width=0.85\linewidth]{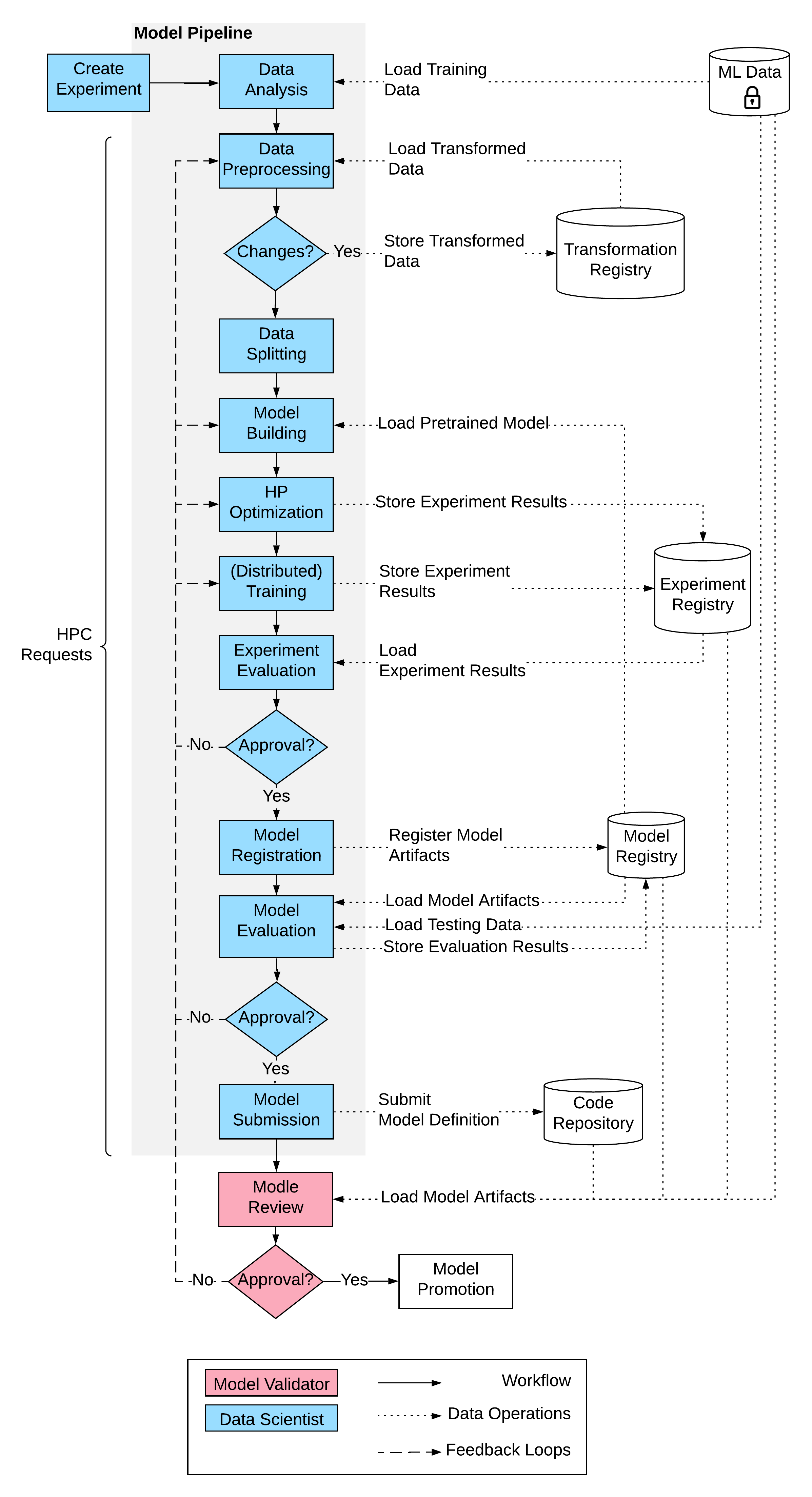}
	\caption{Definition of the abstract model pipeline.}
	\label{fig:model_flow}
\end{figure}

Similar to the \stepDataEngineer{Data Pipeline}, the \stepDataScientist{Model Pipeline} abstracted in the overview flow chart is defined in detail in \cref{fig:model_flow}.
There are two main actors within the model pipeline: the \actorDataScientist{} and the \actorModelValidator{}.
The \actorDataScientist{}'s aim is to analyze and explore data, extract features, and prototype models in an experimental manner~\cite{burkov_machine_2020}.
The \actorModelValidator{} takes responsibility for the project and ensures that the business requirements are fulfilled~\cite{dubovikov_managing_2019}.
Analogously to the \stepDataEngineer{Data Pipeline}, a \actorDataScientist{} can request \ac{hpc} Resources on-demand throughout all steps of the \stepDataScientist{Model Pipeline}.

Initially, an experiment is created by either modifying an existing version of the \entityNeutral{Code Repository} or creating an experiment from scratch.
As a first step of the pipeline, the \actorDataScientist{} analyzes the provided training data, whereby sensitive information may be hidden or masked.
Direct access to the testing data may as well not be granted, \eg, due to privacy reasons.
After the \stepDataScientist{Data Analysis}, the data is preprocessed to be compatible as model input if necessary.
At this step, the \actorDataScientist{} may choose to store a version of the transformed data within the \entityNeutral{Transformation Registry} for faster and more convenient access on further iterations.
If the data transformation changes on subsequent iterations, the data needs to be stored again.
Otherwise, it can simply be loaded into the experimentation environment.
Next, the data is split into a training and validation sets, before the \actorDataScientist{} starts building a model.
In contrast to the \stepDataEngineer{Data Splitting} step of the \stepDataEngineer{Data Pipeline}, this \stepDataScientist{Data Splitting} step further divides the training data and is not relevant to other pipelines.
As opposed to the test dataset, the validation set is not further used.

Upon \stepDataScientist{Model Building}, one has the option to load pre-trained models from the \entityNeutral{Model Registry}.
Within the flow chart in \cref{fig:model_flow}, the \entityNeutral{Model Registry} is illustrated as a single persistence entity.
However, pre-trained models can be loaded from any private or public registry.
Thus, there may be multiple registries available to the \actorDataScientist{} for loading pre-trained models.
As an optional step before training a model, \stepDataScientist{HP Optimization} helps a data scientist finding the optimal configuration of a defined model.
Subsequently, a \stepDataScientist{Training} job is launched, which can optionally be distributed over computing instances, given that the training is a resource-intensive activity.

Whenever a \stepDataScientist{Training} or \stepDataScientist{HP Optimization} job is executed, the corresponding metadata, metrics, and results are stored within the \entityNeutral{Experiment Registry}.
This acts as a central location for experimentation history, not only available to the \actorDataScientist{} building the current model, but also to others for review.
Thereby, the evolution of a model remains comprehensible.

At the step of \stepDataScientist{Experiment Evaluation}, the \actorDataScientist{} review their training experiment based on the validation results and other metrics~\cite{zheng_evaluating_2015}.
If not satisfied, they return to the previous stages of the \stepDataScientist{Model Pipeline}.
Otherwise, the model is registered on the \entityNeutral{Model Registry} to make it available for \stepDataScientist{Model Evaluation}.
At this point, the testing data is loaded together with the model artifacts to test the produced model.
In case the testing data contains sensitive information, the \stepDataScientist{Model Evaluation} can be conducted in a secure environment.
Consequently, the test results are stored to the metrics of the model within the \entityNeutral{Model Registry}.
Based on the evaluation results, the \actorDataScientist{} then decides to either submit the model for review or return to previous steps of the model pipeline.

During the \stepDataScientist{Model Submission} step, the source code of the registered model is submitted to the \entityNeutral{Code Repository}, although the source code is within the model artifacts.
However, the \entityNeutral{Code Repository} should hold all the code needed to reproduce a workflow iteration.

After the \actorDataScientist{} submitted their model, the \actorModelValidator{} reviews the registered and evaluated model by loading corresponding artifacts.
A \stepModelValidator{Review} focuses on model quality and can include various metrics such as accuracy, sensitivity, precision, different error measures, or ranking methods~\cite{shahinfar_how_2020}.
These metrics can be compared to other produced models, possibly already deployed to production.
If the results are not satisfactory, the \actorDataScientist{} can return to specific steps within the \stepDataScientist{Model Pipeline} and improve the model version at the \actorModelValidator{}'s request.
Furthermore, the \actorModelValidator{} can instruct the \actorDataEngineer{} to collect new or more qualitative data, as illustrated in the overview flow chart in \cref{fig:overview_flow}.
In case of approval, the created model is promoted to production, which triggers the \stepDevOpsEngineer{Deployment Pipeline}.

\subsection{Deployment Pipeline}
\label{subsec:deploy_flow}

\begin{figure}[tb]
	\centering
	\includegraphics[width=0.8\linewidth]{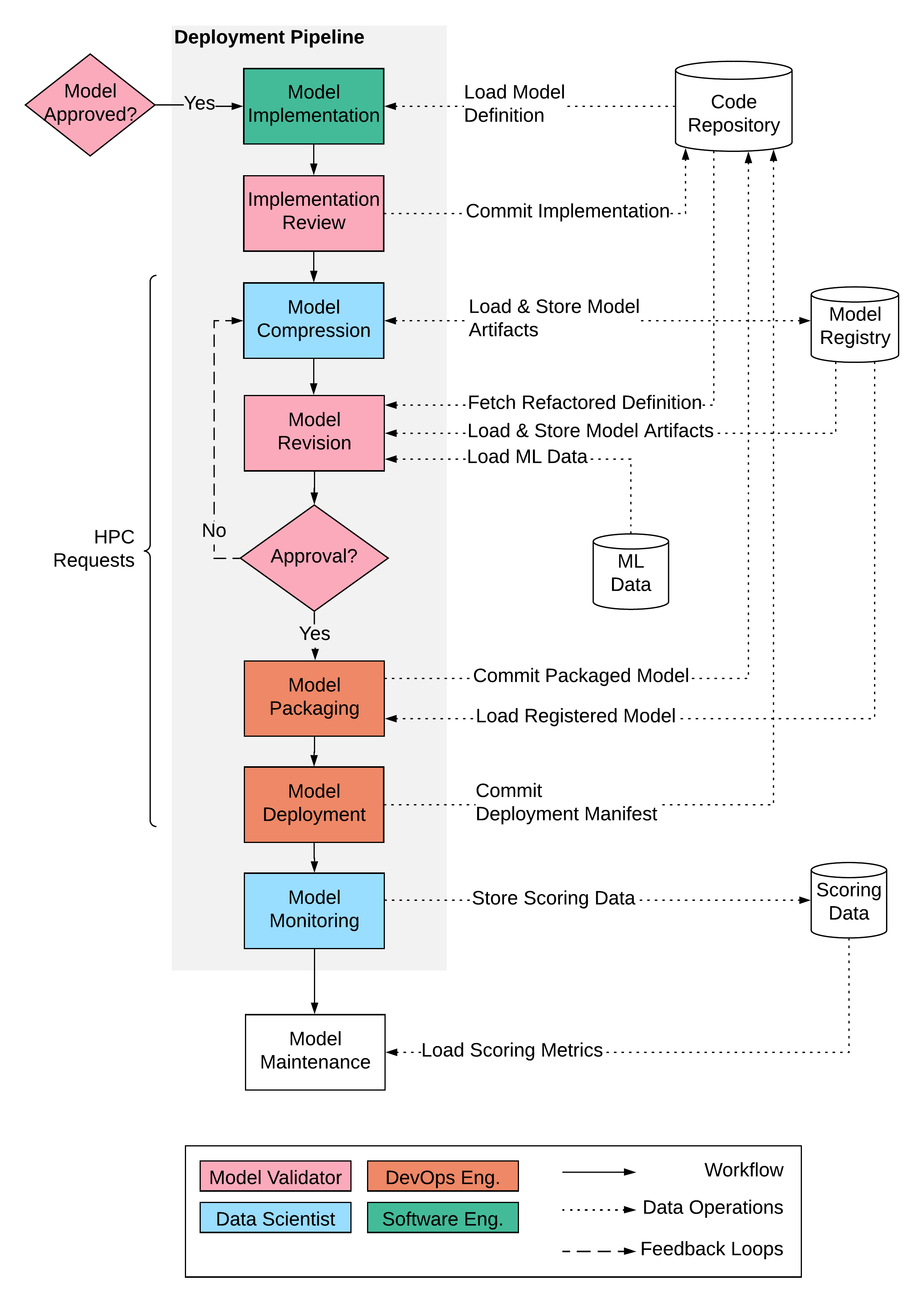}
	\caption{Definition of the abstract deployment pipeline.}
	\label{fig:deploy_flow}
\end{figure}

Once a model has been approved and promoted to production, the goal is to deploy the model.
Besides other roles involved in the \stepDevOpsEngineer{Deployment Pipeline}, the \actorDevOpsEngineer{} is primarily responsible for bringing a model into the deployment environment.
As in the \stepDataEngineer{Data Pipeline} and \stepDataScientist{Model Pipeline}, there are certain steps within the pipeline that require computing resources on-demand, such as \stepDevOpsEngineer{Model Deployment}.
The \stepDevOpsEngineer{Deployment Pipeline} is illustrated in \cref{fig:deploy_flow}.

In case the model is not yet suitable for future retraining, the source code needs to be refactored into a performant, automation, and testing-friendly form.
This task called \stepSoftwareEngineer{Model Implementation} is performed by a \actorSoftwareEngineer{} as a first step of the \stepDevOpsEngineer{Deployment Pipeline}.
A \actorSoftwareEngineer{} has advanced knowledge on runtime performance and memory usage and can therefore make the source code more efficient for retraining~\cite{lakshmanan_machine_2020}.
They further provide expertise in \ac{api} design for effective prediction requests.
The implementation is further \stepModelValidator{Reviewed} by the \actorModelValidator{} and stored into the \entityNeutral{Model Registry} with the corresponding model.

As an optional step within the \stepDevOpsEngineer{Deployment Pipeline}, the model can be \stepDataScientist{compressed} to reduce size and latency, \eg, pruning and quantization, low-rank factorization, convolutional filters, and knowledge distillation.
In this case, the model needs to be tested and compared to the initial model to prevent a significant decrease in accuracy.
For certain model compression techniques, retraining the model is additionally required before testing~\cite{cheng_survey_2020}.
This is conducted during the stage of \stepModelValidator{Model Revision}.
The model artifacts in the \entityNeutral{Model Registry} have to be updated, if the model definition has been refactored.

After the preceding preparation steps, the \actorDevOpsEngineer{} \stepDevOpsEngineer{packages the model} into an appropriate form for inference, which wraps the loaded model with an additional layer to serve prediction requests.
Subsequently, they write a manifest that defines the deployment configuration, and finally \stepDevOpsEngineer{deploy the model} to production.

Once the model is deployed, a \actorDataScientist{} is required to \stepDataScientist{monitor the model} for environment changes.
Thus, input data from prediction requests, together with their metadata and results, are stored in a database for analysis.
In case the model performance declines or other issues occur, the final step of \stepNeutral{Model Maintenance} initializes the next iteration based on the interpretation of the \entityNeutral{Scoring Data}.

\section{Prototype Implementation}
\label{sec:prototype_implementation}

Throughout the following sections, we describe the implementation of a minimum viable prototype that enables the \ac{dl} workflow defined in \cref{sec:workflow}.
Thereby, we demonstrate the transfer of the conceptual workflow into practical implementation.
To do so, we select available technologies that fulfill the requirements given through the characteristics of \ac{dl} listed in \cref{sec:introduction}, and our defined workflow.
Technical instructions on the prototype can be found in the associated \textit{GitHub} repository at \url{\replicationurl}.

\subsection{Technologies of Choice}
\label{subsec:tools_of_choice}

There is a relatively new, but fast-growing market for technologies that partially or completely facilitate the \ac{ml} workflow.
At the time of this paper, the landscape of tools available is broad and not fully mature~\cite{mlops}.
Few are targeted towards \ac{dl} only, as their vendors provide generic solutions to problems of the \ac{ml} workflow.
Nevertheless, these solutions can often be transferred to various subtypes of \ac{ml}.
There are some emerging platforms that try to cover the complete \ac{ml} workflow, such as \textit{Vertex AI} or \textit{Sagemaker} maintained by Google and Amazon, respectively.
However, these solutions can introduce immense costs, especially with frequent usage of \acp{gpu}~\cite{google_pricing, amazon_pricing}.
On behalf of reproducibility, we restrict our selection of tools to being deployable on-premise, available for free, and preferably open-source.

\subsubsection{Hardware and Infrastructure}

We used the computing and storage resources provided by the ScienceCloud~\cite{sciencecloud} of the \ac{uzh}, which lets us provision \acp{vm}:
\begin{inparaenum}[(1)]
	\item a large \ac{vm} with \num{32} \acp{vcpu}, a single Nvidia Tesla T4 \ac{gpu} and \SI{128}{\giga\byte} of \ac{ram};
    \item a small \ac{vm} with \num{2} \acp{vcpu} and \SI{8}{\giga\byte} of \ac{ram}.
\end{inparaenum}

The specifications of the large \ac{vm} are chosen regarding the minimal resources to host all technologies required to run the \ac{dl} workflow and meet the demands described in \cref{sec:introduction}.
A single \ac{gpu} is the maximum amount available per \ac{vm} on the ScienceCloud, but to demonstrate distributed training, at least two \acp{gpu} would be required.
To independently and securely host a \entityNeutral{Code Repository}, a second \ac{vm} was introduced with a near minimal specification to spare resources.
Both \acp{vm} run \textit{Ubuntu} 18.04 as their operating system.

To meet the requirement of scalability within the \ac{dl} workflow, we selected \textit{Kubernetes} as our infrastructure of choice.
It serves as an open-source system for scalable container orchestration~\cite{kubernetes}.
In the context of this paper, we use \textit{Microk8s}, a lightweight upstream \textit{Kubernetes} with low operation costs and \ac{gpu} support~\cite{microk8s}.
However, the prototype is portable to any \textit{Kubernetes} version.
For our use case, a \textit{MicroK8s} single-node cluster is hosted on the large \ac{vm}.
As a container technology of choice, we use \textit{Docker} due to being widely used across the developer community.
Furthermore, \textit{DockerHub} serves as the container registry to push and pull images.

\subsubsection{Workflow Tools}
\label{subsec:workflow_tools}

% \begin{table}[tb]
% 	\caption{Technologies used to implement the \ac{dl} workflow, not including the underlying infrastructure}
% 	\label{tab:workflow_technologies}
% 	\centering
% 	\resizebox{1.0\linewidth}{!}{
% 	\input{tables/workflow_technologies}
% 	}
% \end{table}

Hereinafter, the technologies that directly support the implementation of the \stepNeutral{workflow steps} and \entityNeutral{persistence entities} are described.
Our selection is based on the recommendations of the MLOps Community~\cite{mlops_community} and Visengeriyeva \etal~\cite{mlops}.
% A summary of the technologies used, their purpose, and what \entityNeutral{persistence entity} they represent can be found in \cref{tab:workflow_technologies}.
It is important to note that this selection is not fixed and could be swapped with any tools with similar features.

We choose \textit{GitLab}~\cite{gitlab} as our \ac{vcs}, representing the \entityNeutral{Code Repository}.
\textit{GitLab} can be hosted on premise, provides mature features for \ac{ci}/\ac{cd} and integrates with \textit{Kubernetes}.

To achieve data lineage, \ie, \stepDataEngineer{Data Pipeline} and version-controlled datasets, \pach~\cite{pachyderm} is selected.
It allows us to execute containerized tasks on a \textit{Kubernetes} cluster in a scalable, parallel, and distributed manner.
\pach can serve as a general object storage technology, thus be used to store unstructured datasets, and as the \entityNeutral{Transformation Registry} by the \actorDataScientist{} to cache transformed data.

We used \textit{Label Studio}~\cite{labelstud} to integrate \stepDataLabeler{Data Labeling} into our workflow implementation.
It is compatible with various types of data, especially unstructured data commonly used in \ac{dl} applications such as computer vision, natural language processing, and audio processing~\cite{amazon_labeling}.
The labels, together with their metadata, are stored to a \textit{PostgreSQL}~\cite{postgresql} database for fast and convenient queries.

For tasks related to the \stepDataScientist{Model Pipeline}, we used a cloud-native platform specifically targeted towards \ac{dl} called \textit{Determined}~\cite{determinedai}.
This platform addresses the need for distributed \stepDataScientist{Training}, \stepDataScientist{HP Optimization}, and compute resource management.
\detai automatically tracks experiments for analysis and additionally provides a \entityNeutral{Model Registry} to store model artifacts.
Thereby, a \actorDataScientist{} can focus on building and optimizing a model.
Under the hood, a \textit{PostgreSQL} instance represents the \entityNeutral{Experiment Registry}, whereas a \textit{MinIO}~\cite{minio} bucket is configured to store artifacts of the \entityNeutral{Model Registry}.
\textit{MinIO} is a widely used, cloud-native object-store.

To \stepDevOpsEngineer{deploy models} at scale to \textit{Kubernetes}, \textit{Seldon}~\cite{seldon} was used.
It supports a large spectrum of \ac{ml} libraries and deployment configurations.
A model can be brought to production by simply building a language wrapper around the model and specifying the container environment.
Although alternative technologies to deploy \ac{ml} models on \textit{Kubernetes}, \textit{Seldon} appeared to be the most mature solution.

We did not implement the \entityNeutral{Scoring Data} persistence entity, as \stepDataScientist{Model Monitoring} would exceed the scope of this paper.
However, a \textit{PostgreSQL} database holding results of requests, and possibly references to provided files stored to \pach, would be applicable.

\subsection{Mapping Abstraction and Implementation}

With the technologies selected in \cref{subsec:workflow_tools}, we can build a \ac{dl} system that implements our abstract workflow of \cref{sec:workflow}.
By mapping the technologies to tasks and \entityNeutral{persistence entities}, we demonstrate how these integrate into the \ac{dl} workflow.
For each pipeline, we will walk through the practical utilization of the technologies.

\subsubsection{Data Pipeline Implementation}
\label{subsubsec:data_pipeline_implementation}

% \begin{figure}[tb]
% 	\centering
% 	\includegraphics[width=1.0\linewidth]{figures/workflow/prototype_data}
% 	\caption{Implementation of the data pipeline, with the selected technologies mapped to the flow chart.}
% 	\label{fig:prototype_data}
% \end{figure}

Besides other tools for \ac{ci}/ac{cd}, the main technologies within the \stepDataEngineer{Data Pipeline} are \pach and \textit{Label Studio}.
%\cref{fig:prototype_data} illustrates how these technologies are integrated into the data pipeline.
The \actorDataEngineer{} defines all steps of the pipeline with a programming language of choice, from \stepDataEngineer{Data Collection} to \stepDataEngineer{Data Validation}.
As mentioned in \cref{subsec:data_flow}, the steps \stepDataEngineer{Data Collection} and \stepDataEngineer{Data Splitting} are not necessarily part of the automated pipeline.
In our case, the \stepDataLabeler{Data Labelling} also remains a manual step.
The \actorDataEngineer{} can define different sources for training and testing data, which implicitly splits the data, and then build separate pipelines.
However, it is important that both data sets are processed the same way, \ie, the same scripts for each step are used.
Otherwise, the datasets could exhibit different characteristics, for example when the training and testing data are validated differently.

Once all steps are defined, the \actorDataEngineer{} packages the scripts into a pipeline by writing a manifest complying to the \textit{Pachyderm} format, which has either \ac{json} or \ac{yaml} format.
Within this manifest, they optionally specify the resources to be used at each step, such as \ac{gpu}, \ac{cpu} and memory.
Additionally, one can define how ingested data is processed, \eg, as streams or in batches.
The \actorDataEngineer{} further defines the \textit{Docker} container, wherein the pipeline is executed.
They then commit their work to the \textit{GitLab} \entityNeutral{Code Repository}.
This triggers the build of the \textit{Docker} image and subsequently a push to \textit{DockerHub}.
Moreover, the pipeline is indirectly deployed to \textit{Kubernetes} via \pach.
The execution of the pipeline is initiated each time data is ingested.

\pach presents input and output repositories for each pipeline.
Thus, the output of the \stepDataEngineer{Data Validation} can automatically be ingested into \textit{Label Studio}.
Once the data has been annotated by the data labeler, the labels are exported in a format of choice into another \pach repository.
From there, the labels are stored to a \textit{PostgreSQL} database, where the testing and training data reside in separate tables.
Within a table, a row keeps information about the label, the corresponding file path to the validated output repository, and metadata, \eg, labeler, date, and dataset version.

\stepDataEngineer{Data Versioning} is automatically performed by \pach in a git-like manner.
Each repository, \ie, bucket, allows branching and annotates ingested data with commit IDs.
Upon further iterations of the \stepDataScientist{Data Pipeline}, a \actorDataEngineer{} can ingest on a new branch or distinguish between dataset versions using the commit ID within the same branch.
Similarly, the pipeline manifest is versioned as well, thereby it remains fully comprehensible how a specific dataset version was produced.
Data lineage is therefore guaranteed.

\subsubsection{Model Pipeline Implementation}

% \begin{figure}[tb]
% 	\centering
% 	\includegraphics[width=1.0\linewidth]{figures/workflow/prototype_model}
% 	\caption{Implementation of the model pipeline, with the selected technologies mapped to the flow chart.}
% 	\label{fig:prototype_model}
% \end{figure}

Within the implementation of the \stepDataScientist{Model Pipeline}, a \actorDataScientist{} mainly interacts with \detai and \pach.
%as illustrated in \cref{fig:prototype_model}.
To initiate an experiment, a new branch within \gl is created, either from an existing branch or from scratch.
While operating locally, they can choose to work in a notebook environment offered by \detai or directly write scripts as their source code.
However, a notebook will have to be downloaded manually and checked into version control.

At the first step of the \stepDataScientist{Model Pipeline}, the training data is loaded from \pach, analyzed, and preprocessed.
The transformed data can be stored to a \pach repository and accessed on further iterations for faster development.
After splitting the data into a train and validation set, they start \stepDataScientist{building the model} using a \detai compatible definition.
Therefore, a trial class must be built that implements a predefined set of member functions for initialization, training, evaluation, and data loading.
Through these restrictions, a \actorDataScientist{} does not need to take care of logging and visualizing metrics, or saving model checkpoints.
Within the process of \stepDataScientist{Model Building}, pre-trained models can be loaded from the internal \detai \entityNeutral{Model Registry} or any external model registry, such as the Hugging Face transformer library~\cite{wolf_transformers_2020}.

Besides a model definition, \detai additionally requires a configuration file in \ac{yaml} format, which specifies \acp{hp}, resource requests, data source, and version, etc.
The \actorDataScientist{} can then use the same model definition with different configuration files for \stepDataScientist{HP Optimization} and \stepDataScientist{(Distributed) Training}.
Moreover, \detai executes jobs on agents within containers scheduled by a master.
A \actorDataScientist{} can either specify software dependencies within a startup script, or build a container image for the agent to run on.
If a \textit{Docker} image is used, it is built and pushed upon a commit to \textit{GitLab} and subsequently pulled by \detai on a run execution.
Within the process of \stepDataScientist{HP Optimization} and \stepDataScientist{(Distributed) Training}, a \actorDataScientist{} can review and compare their executed jobs on the \detai \ac{ui} until a satisfying model.

Once a \actorDataScientist{} approves of the validated model, they can \stepDataScientist{register} a selected model checkpoint through the \detai \ac{cli}.
To synchronize the \entityNeutral{Code Repository} with the latest model version, the corresponding model definition has to be downloaded manually from the \entityNeutral{Model Registry} before committing.
At the step of \stepDataScientist{Model Evaluation}, a commit to the experiment branch on \textit{GitHub} triggers the model testing, which loads the testing data from \pach and the latest model version from the \entityNeutral{Model Registry} for \stepDataScientist{evaluation}.
This commit is required as a model should be evaluated remotely and \gl cannot listen for changes in the \detai registry due to a lack of change events.
The results are then written to the model metrics and additionally presented as a \textit{GitLab} pipeline artifact to the whole team.

If a \actorDataScientist{} agrees, they perform a \stepDataScientist{Model Submission} with a merge request on \textit{GitLab}.
The \actorModelValidator{} then reviews the experiment and, if approved, the \stepDevOpsEngineer{Deployment Pipeline} is triggered.

\subsubsection{Deployment Pipeline Implementation}

After the \actorModelValidator{} has approved a model, it is prepared for production.
%An illustration of the implemented deployment pipeline is available in \cref{fig:prototype_deploy}.
In the context of our minimum viable prototype, we skip two steps of the \stepDevOpsEngineer{Deployment Pipeline}:
\begin{inparaenum}
	\item we do not perform \stepDevOpsEngineer{Model Compression}, as we deploy to \textit{Kubernetes} and inference latency is neglected;
	\item \stepDevOpsEngineer{Model Monitoring} is omitted as this would exceed the scope of this paper.
\end{inparaenum}

First, a \actorSoftwareEngineer{} fetches the model definition from \textit{GitLab} to refactor the source code, which also includes the sources for \stepDataScientist{Data Preprocessing}.
After the \stepModelValidator{Implementation Review} and \stepDataScientist{Model Compression}, we retrain, evaluate and test the model for performance, \eg, model size and inference latency, at the step of \stepModelValidator{Model Revision}.
Therefore, the \actorModelValidator{} loads model artifacts from \detai, the refactored model definition from the \entityNeutral{Code Repository}, and the datasets from \pach.

Once the revision is approved, the \actorDevOpsEngineer{} builds the model wrapper for \textit{Seldon}.
The wrapper is essentially a Python class that defines at minimum how the model is loaded and how inputs are preprocessed for prediction.
Additionally, a \textit{Docker} image defines the container for the model environment at run-time.
At the step of \stepDevOpsEngineer{Model Deployment}, the \actorDevOpsEngineer{} commits the deployment manifest to the main branch  to trigger the deployment.
The deployment manifest specifies what resources are available to the model.
Then, a \textit{GitLab} pipeline builds and pushes the \textit{Docker} image, and deploys the packaged model to \textit{Kubernetes} using \textit{Seldon}.

% \begin{figure}[tb]
% 	\centering
% 	\includegraphics[width=1.0\linewidth]{figures/workflow/prototype_deploy}
% 	\caption{Implementation of the deployment pipeline, with the selected technologies mapped to the flow chart.}
% 	\label{fig:prototype_deploy}
% \end{figure}

\section{Use Cases}
\label{sec:use_cases}

To investigate the practicability of our implementation and the embedded pipelines established in \cref{sec:prototype_implementation}, we apply two use cases to the prototype.
These use cases address common but distinguished problems of \ac{dl}, using different frameworks to provide variety.
The source code for each use case including the setup is available on \textit{GitHub} at \url{\replicationurl}.

\subsection{News Classification}

In our first use case, the goal is to address an \ac{nlp} problem.
Therefore, a multinomial news classification model trained and evaluated on the BBC datasets~\cite{greene_practical_2006} is constructed using \textit{PyTorch}~\cite{paszke_pytorch_2019}.

The BBC datasets consists of \num{2225} text documents that represent news articles from the years 2004 and 2005, collected from the official BBC news website~\cite{greene_practical_2006}.
The documents in English have various lengths and are divided into the following categories: business, entertainment, politics, sport, and tech.
% The datasets are available for download in two versions~\cite{Greene2020}: a preprocessed version, which includes stemming, stop-word removal and low term frequency count, and a raw version which provides the unprocessed articles.
To be able to demonstrate the complete data pipeline, we use the raw version.

As a \actorDevOpsEngineer{}, we initially set up a \gl repository with a main folder and a \gl pipeline for \ac{ci}/ac{cd}.
The main folder includes subdirectories for data, model, deployment, and test code.
The \entityNeutral{Code Repository} has two branches, one for development (dev) and one for production (main).

\paragraph{Data Pipeline}
% A visualization of the data pipeline corresponding to this use case can be found in \cref{fig:bbc_data}.
We assume the role of a \actorDataEngineer{} and perform the first steps of the \stepDataEngineer{Data Pipeline} manually using Python scripts.
By downloading the dataset locally, giving each file a unique ID, merge each category and randomize the order, we prepare the data set for ingestion.
Then, the dataset is split into a train and test set, \ie, hold out set, using an \numrange[range-phrase = --]{80}{20} ratio~\cite{burkov_machine_2020}
% Undoubtedly, there would be more sophisticated approaches~\cite{Lakshmanan2020}.
and we automatically create labels in \textit{LabelStudio} \ac{json} format for each article, as we do not possess the resources for manual labeling.

% \begin{figure}[tb]
% 	\centering
% 	\includegraphics[width=1.0\linewidth]{figures/usecase/bbc_data}
% 	\caption{The data pipeline of news classification use case.}
% 	\label{fig:bbc_data}
% \end{figure}

A simple Python script then defines how data is ingested, cleaned, and validated.
We replace characters not being UTF-8 conform, ensure that the articles have a minimum character length and TXT file format.
Articles not corresponding to the minimum character length or file format are invalid and therefore discarded.
Valid articles are copied to the output repository defined by \pach.
After the steps of \stepDataEngineer{Data Ingestion}, \stepDataEngineer{Data Cleaning} and \stepDataEngineer{Data Validation} are defined, we initialize two separate \pach repositories for the train and test datasets.
Two pipelines for each dataset are then constructed, which essentially execute the same afore-mentioned script with different input paths
% , as illustrated in \cref{fig:bbc_data}.
% For demonstration purposes, the validation pipeline requests one \ac{cpu} and processes ingested data in batches.

We then prepare the \textit{Docker} container for both pipelines to run on.
The image installs the required packages and pulls the source code defining the pipeline.
In this use case, the pipeline is directly committed to the production branch in the \textit{GitLab} repository.
% Certainly, a more realistic application would require review and testing of the pipeline.
\textit{GitLab} then automatically builds and pushes the \textit{Docker} image and deploys the pipelines to \pach on \textit{Kubernetes}.

Subsequently, both training and testing data can be ingested into the corresponding pipeline with a single \pach \ac{cli} command.
The validated data is synchronized into \textit{Label Studio} and available for manual labeling.
An additional \pach pipeline facilitates the export of the labels from \textit{Label Studio} into the respective table within the \textit{PostgreSQL} database.
A single row holds the label itself, the file path to the article in the \pach repository, additional metadata about the label, and the matching \pach branch for data lineage.
Note that in this use case, a model is not automatically retrained upon ingestion of new data or changes to the pipeline, but this could be enabled using an additional \textit{GitLab} pipeline stage.

\paragraph{Model Pipeline}
% \label{subsubsec:use_case_1_model_pipeline}
We take over the role of a \actorDataScientist{} and start the first step of \stepDataScientist{Create Experiment}.
We create a new branch within the \textit{GitLab} repository, in this case from the main branch.
As we are already provided with sufficient knowledge about the BBC dataset, the step of \stepDataScientist{Data Analysis} is omitted.
Nevertheless, the labels that reside in the \textit{PostgreSQL} training table are loaded together with the corresponding news articles from \pach.

To transform the text data into input conform to the model, a simple \ac{nlp} approach is applied.
This includes Porter stemming~\cite{porter_algorithm_2006} and token count vectorization using the feature extraction capabilities of \textit{scikit-learn}~\cite{pedregosa_scikitlearn_2011}.
As a vocabulary, we use the one provided by Greene~\cite{insight_bbc}.
The target classes are similarly encoded into numeric values.
The preprocessed data is then again split into a training and validation sets at an \numrange[range-phrase = --]{80}{20} ratio~\cite{burkov_machine_2020}.
In this use case, the preprocessed data is not stored to a \entityNeutral{Transformation Registry} for simplicity.
% However, the data frame could certainly be stored and versioned in a \pach repository.

During \stepDataScientist{Model Building}, we locally create the required trial class for \detai using \textit{PyTorch}.
Our basic model is a \acl{nn} of five linear layers with layer normalization and dropout applied.
Additionally, a separate configuration file defines the \pach repository and branch, hyperparameters such as dropout rate, hidden layer size, learning rate, and metadata about the experiment.

To start the \stepDataScientist{HP Optimization}, we specify a reasonable search space for each hyperparameter and then submit our configuration to the cluster using the \detai \ac{cli}.
Vocabulary, classes, and a list of required software packages are automatically uploaded to the \entityNeutral{Model Registry}.
In the context of this use case, we use the \ac{asha} algorithm, which supports early stopping of low performing configurations~\cite{li_massively_2020}.
\detai then presents various visualizations and metrics to find the optimal hyperparameter configuration.
Since the parameters in the configuration are loaded at run-time, we can consequently use multiple configuration files for \stepDataScientist{HP Optimization} and \stepDataScientist{(Distributed) Training} with the same code for preprocessing and model definition.
Thus, we write an additional configuration file for training using the results of \stepDataScientist{HP Optimization}, and again submit everything to the cluster.
% As our infrastructure only compromises one \ac{gpu}, we do not demonstrate distributed training.
% However, \detai enables networking, data loading and fault tolerance with the specification of a single configuration argument.

When finally arriving at a satisfying performance, we select the checkpoint UUID of the preferred experiment, register the model through the \detai \ac{cli}, and push our local changes to the remote repository to trigger the \gl pipeline and evaluate our model.
\stepDataScientist{Model Evaluation} is facilitated through a Python unit test, which loads the model and the test data and only passes if a minimum accuracy threshold of \num{0.7} is reached.
% This value is set arbitrarily and is usually dependent on the project requirements.
Within an initial iteration of the \ac{dl} workflow, either the \actorDataScientist{} himself or a \actorDevOpsEngineer{} writes test cases.
Who is responsible for testing a model highly depends on whether the test data is confidential.

Subsequently, we can request a merge onto the development branch.
The \gl pipeline produces a text document with the test results and a list of all model versions with their corresponding checkpoint UUID.
Thereby, as the \actorModelValidator{}, we can conclude the experiment within the \detai \ac{ui}.
If the model is adequate for production, the merge request is accepted.

\paragraph{Deployment Pipeline}
% \label{subsubsec:use_case_1_deploy_pipeline}
The experiment is now approved and merged into the development branch.
As this is only a demonstration use case, we do not consider model performance, and the steps of \stepSoftwareEngineer{Model Implementation}, \stepModelValidator{Implementation Review}, and \stepModelValidator{Model Revision} are omitted.

We further assume the role of a \actorDevOpsEngineer{}.
To package the model for deployment, we build a Python class that has two functions: one function initializes the model, therefore loads the model including its artifacts by downloading from the \entityNeutral{Model Registry} within \detai.
% This includes the vocabulary and classes used during training and the registered model checkpoint.
The other function defines how a prediction is performed.
In this context, the input is transformed the same way as during model \stepDataScientist{Training}, \ie, stemmed and tokenized using the vocabulary.
The numerical output returned from the model is resolved to the respective category using the label encoding.

To deploy a model with \textit{Seldon}, we define a \textit{Docker} image that copies the model wrapper and installs the required dependencies.
The deployment manifest is specified using a \ac{yaml} file, serving the packaged model as a \ac{rest} \ac{api} and specifying the Docker image as the surrounding container.
% At this stage, we could additionally specify resource requests such as the number of \acp{gpu} to be used for model serving.
% However, as only one \ac{gpu} is available, this is omitted.

For continuous delivery of future model improvements, the steps of building and pushing the \textit{Docker} images as well as deploying to \textit{Kubernetes} are automated through \gl pipeline jobs.
These are specified to be executed only on pushes to the production branch.
In the context of testing, the model could additionally be deployed to a development environment as an additional staging process.

\subsection{Fashion Classification}

The second use case derives a multinomial image classification model using \textit{TensorFlow}~\cite{abadi_tensorflow_2015} and the Fashion-MNIST dataset provided by Zalando Research~\cite{xiao_fashionmnist_2017}.
Thereby, a classical computer vision example of \ac{dl} is tackled.

The Fashion-MNIST dataset consists of \num{60000} training and \num{10000} testing images of clothing articles~\cite{xiao_fashionmnist_2017}.
The grayscale $28 \times 28$ images are divided into ten categories: T-shirt/top, trouser, pullover, dress, coat, sandal, shirt, sneaker, bag ,and ankle boot.
The dataset is available for download on \textit{GitHub}~\cite{zalando_fashion} as separate binary files representing labels and images.

As in the first use case, we set up a \gl repository, including a pipeline and main folder containing the sources in subdirectories, and starting with two branches for development and production.

\paragraph{Data Pipeline}
% As we already demonstrate a fully implemented data pipeline with unstructured data in our first use case, the approach in this second use case is simplified.
The pipeline resembles our first use case in many aspects, but instead of storing text documents, we would store image files in the \pach repositories.
To simplify this use case, we directly store the compressed binary files containing the preprocessed fashion images and labels in a \pach repository.
We assume that the data has already been cleaned and validated, thus no \stepDataEngineer{Data Pipeline} is constructed.

\paragraph{Model Pipeline}
Similar to the first use case, initially an experiment branch is created from the production branch.
Assuming the role of a \actorDataScientist{}, we load the labels and source images from the respective \pach repository and decompress them.
% Again, the step of \enquote{Data Analysis} is omitted as sufficient knowledge about the dataset is already provided.
To preprocess the dataset, the pixel values within a scale of \numrange{0}{255} are normalized to a scale of \numrange{0}{1}.
Similar to the first use case, we do not store the preprocessed data frame into a \entityNeutral{Transformation Registry}, although \pach could very well be used therefor.
A validation set is then split off at an \numrange[range-phrase = --]{80}{20} ratio.

To demonstrate the application of different \ac{ml} frameworks, \textit{TensorFlow Keras} is used to build our image classification model, based on an example provided in the official documentation~\cite{tensorflow_tutorial}.
For \stepDataScientist{Model Building}, a sequential model is constructed composed of a flat input layer, a dense hidden layer of variable size, and a dense layer fixed at the number of output categories.
The remaining steps follow the process of the first use case.
% : we optimize the hyperparameters, train the model, review experiments, register a model, evaluate it and finally request a merge to the development branch in the \gl code repository.

\paragraph{Deployment Pipeline}
As in the first use case, we omit the first four steps of the deployment pipeline.
Again, as \actorDevOpsEngineer{}, we build a Python class that packages, \ie, wraps the model.
At this step, only the classes defined in the \stepDataScientist{Model Pipeline} are loaded as model dependencies.
We then construct a \textit{Docker} image that starts the \textit{Seldon} microservice, and serves the model as a \ac{rest} \ac{api}.
Merging the source code from the development branch onto the main production branch ultimately triggers the build of the \textit{Docker} image and finally the deployment to \textit{Kubernetes}.

\section{Conclusions and Future Work}
\label{sec:conclusions}

In this paper, we described a detailed step-by-step \ac{dl} workflow, split into three pipelines, such that each component is independently reproducible and thereby enabling fast iterations.
We provide a general guideline on the \ac{dl} development process that helps to build \ac{dl} models and continuously improve them through efficient and reproducible iterations.
Additionally, our recommended set of technologies can be used as a reference for future implementations.
% These pipelines in turn can further include sequences of automated steps.
% For example, the data management pipelines constructed in \cref{subsubsec:data_pipeline_implementation} include separated downstream pipelines for preparing data and storing labels, respectively.
% Together with additional pipelines for \ac{ci}/ac{cd}, a project team can quickly face the challenge of \enquote{pipeline jungles} introduced by Sculley \etal~\cite{sculley_hidden_2015}.
% Managing these pipelines thus requires frequent knowledge exchange, versioning, and well-connected interfaces.

Considering our abstraction of the \ac{dl} workflow, it becomes apparent that there is a dependency on various persistence entities with different functions.
This increases data management costs and complicates collaboration, as knowledge about storing and retrieving data is required.
On the contrary, these \entityNeutral{persistence entities} are required to keep the workflow reproducible.
With continuous iterations of the \ac{dl} lifecycle, it becomes important to align the related data between \entityNeutral{persistence entities}.
As an example, additional effort is required to keep the registered model and the version in the \entityNeutral{Code Repository} aligned.
From a technical perspective, both the \entityNeutral{Model Registry} and \entityNeutral{Code Repository}, are crucial.
On the one hand, \acp{vcs} do not allow storing large artifacts such as a \ac{dl} model.
On the other hand, a \entityNeutral{Model Registry} cannot provide a history of code changes, and efficiently hold all source code of the \ac{dl} workflow.
An adopted \ac{vcs} concept that manages \ac{ml} data, model artifacts, workflow source code, etc., as a single set of dependencies in a central location would facilitate the \ac{dl}, and the \ac{ml} workflow in general.
% There are certain tools such as \textit{DVC} and \textit{CML}~\cite{Iterative2021} that focus on this concept.
% However, these are incompatible with the requirements of the \ac{dl} workflow since they do not provide scalability and compute resource orchestration.

Through a prototype, we showed that it is possible to translate the conceptual idea to practice using the latest technologies available on the market.
It becomes clear that a large amount of different technologies is necessary to execute the complete \ac{dl} lifecycle.
Although there are all-in-one solutions available, these technologies often suffer from vendor lock-in and are associated with high costs, and they are not directly targeted towards \ac{dl}.
The large technology stack imposes the need for interfaces between tools, activities, and roles.
This in turn introduces additional hazards to the implemented workflow. For example, we use a complex constellation of technologies within the data pipeline, which can become confusing and error-prone.
% Taking \stepDataLabeler{Data Labeling} as an example: we use \pach since it allows building scalable data pipelines, and managing compute resources.
% As \pach does not provide any labeling features, \textit{Label Studio} is introduced.
% To store our labels to a \textit{PostgreSQL} table, \pach again facilitates export and import.
% Thus, we use a complex constellation of technologies within the data pipeline, which can become confusing and error-prone.

With the application of two distinguished use cases, we demonstrated the practicability of our \ac{dl} workflow.
However, the use cases do not represent the complexity of real-world challenges.
Future work should therefore focus on evaluating the proposed concept in various industries to find possible alterations or inconsistencies.
For instance, a field study could compare the proposed workflow to the processes within companies that have already brought \ac{dl} into use.
% To do so, one could select a set of different companies, possibly with varying cultures, and analyze their workflow with the associated data science team.
By accompanying multiple projects and workflow iterations, all performed activities are collected and mapped to our components. 
% to see whether each step is actually present and performed by the defined role.
% Interviews with the involved developers could protocol the reasons for their actions.
This procedure would ultimately highlight abundant or missing steps within the proposed workflow, inconsistencies in liabilities or more suitable technologies.
% Moreover, this field study would reveal inconsistencies in our suggested roles and liabilities.
% Such an evaluation could additionally lead to the findings of more applicable technologies for an improved prototype.
% On the other hand, further research could address the development of new tools to overcome the limitations of the current tool landscape.
% For example, managing model-related data, \ie, training and testing data, large model artifacts, related source code, etc.
% as a coherent set is still a major challenge.
% From a broader perspective, an all-in-one solution for the \ac{dl} workflow could resolve the limitations of loosing context due to the high number of interfaces.

\balance
\bibliography{references, urls}

\end{document}